\documentclass[prb,reprint,twocolumn,superscriptaddress,citeautoscript,showpacs,amsmath,amssymb,floatfix]{revtex4-2}
\usepackage{graphicx}
\usepackage{amsmath}
\usepackage{amssymb}
\usepackage{mathtools}
\usepackage{textcomp,gensymb} 
\usepackage{bm} 
\usepackage{float}
\usepackage{setspace}
\usepackage{dcolumn}
\usepackage{array}
\newcommand{\minus}{\scalebox{0.75}[1.0]{$-$}}

\usepackage{changepage}
\usepackage{tikz}
\usepackage{contour}
\usepackage[left]{lineno}
\usepackage{blindtext}
\usepackage{setspace}
\usepackage[none]{hyphenat}
\usepackage[colorlinks=true,citecolor=blue,linkcolor=blue,pdftex,hypertexnames=false,urlcolor=blue,linktocpage=true]{hyperref}
\usepackage{color}
\def\newr{\color{black}}

\def\beps{\mbox{\boldmath$\epsilon$\unboldmath}}
\definecolor{ashgray}{rgb}{0.7,0.75,0.71}
\definecolor{mspringgreen}{rgb}{0, 0.8, 0.1}
\definecolor{auburn}{rgb}{0.43, 0.21, 0.1}
\definecolor{ao(english)}{rgb}{0.0, 0.5, 0.0}
\definecolor{afw}{rgb}{0.95, 0.95, 0.96}
\definecolor{magnolia}{rgb}{0.97, 0.96, 1.0}
\definecolor{wsmk}{rgb}{0.96, 0.96, 0.96}

\newcommand{\lco}{La$_2$CuO$_{4}$}
\newcommand{\lsco}{La$_{2-x}$Sr$_x$CuO$_{4}$}
\newcommand{\lbco}{La$_{2-x}$Ba$_x$CuO$_{4}$}

\newcolumntype{M}[1]{>{\centering\arraybackslash}m{#1}}

\usepackage{dcolumn}

\newcolumntype{d}[1]{D{.}{.}{#1}}

\begin{document}
\title{Reinvestigation of crystal symmetry and fluctuations in La$_2$CuO$_4$}
\author{A. Sapkota}
\affiliation {Condensed Matter Physics \&\ Materials Science Division, Brookhaven National Laboratory, Upton, New York 11973-5000, USA}
\author{T. C. Sterling}
\affiliation {Department of Physics, University of Colorado at Boulder, Boulder, Colorado 80309, USA}
\author{P. M. Lozano}
\affiliation {Condensed Matter Physics \&\ Materials Science Division, Brookhaven National Laboratory, Upton, New York 11973-5000, USA}
\affiliation{Department of Physics and Astronomy, Stony Brook University, Stony Brook, New York 11794-3800, USA}
\author{Yangmu Li}
\affiliation {Condensed Matter Physics \&\ Materials Science Division, Brookhaven National Laboratory, Upton, New York 11973-5000, USA}
\author{Huibo Cao}
\author{V. O. Garlea}
\affiliation{Neutron Scattering Division, Oak Ridge National Laboratory, Oak Ridge, Tennessee 37831, USA}
\author{D. Reznik}
\affiliation {Department of Physics, University of Colorado at Boulder, Boulder, Colorado 80309, USA}
\affiliation{Center for Experiments on Quantum Materials, University of Colorado at Boulder, Boulder, Colorado 80309, USA}
\author{Qiang Li}
\affiliation {Condensed Matter Physics \&\ Materials Science Division, Brookhaven National Laboratory, Upton, New York 11973-5000, USA}
\affiliation{Department of Physics and Astronomy, Stony Brook University, Stony Brook, New York 11794-3800, USA}
\author{I. A. Zaliznyak}
\author{G. D. Gu}
\author{J. M. Tranquada}
\affiliation {Condensed Matter Physics \&\ Materials Science Division, Brookhaven National Laboratory, Upton, New York 11973-5000, USA}

\date{\today}
\begin{abstract}
New surprises continue to be revealed about La$_2$CuO$_4$, the parent compound of the original cuprate superconductor.  Here we present neutron scattering evidence that the structural symmetry is lower than commonly assumed.  The static distortion results in anisotropic Cu-O bonds within the CuO$_2$ planes; such anisotropy is relevant to pinning charge stripes in hole-doped samples.  Associated with the extra structural modulation is a soft phonon mode. If this phonon were to soften completely, the resulting change in CuO$_6$ octahedral tilts would lead to weak ferromagnetism.  Hence, we suggest that this mode may be the ``chiral'' phonon inferred from recent studies of the thermal Hall effect.  We also note the absence of interaction between the antiferromagnetic spin waves and low-energy optical phonons, in contrast to what is observed in hole-doped samples. 
\end{abstract}

\maketitle

\section{Introduction \label{Intro}}

Just as the discovery of high-temperature superconductivity in \lbco\ \cite{bedn86,taka87,bedn87} led to a continuous process of revelation in doped cuprates \cite{keim15}, a similar evolution of understanding has taken place with the parent compound \lco\ (LCO).  In an earlier burst of enthusiasm for transition-metal oxide compounds, it had been determined by X-ray diffraction that LCO transforms from a high-temperature tetragonal (HTT) phase (then known as the K$_2$NiF$_4$-type structure) to a low-temperature orthorhombic (LTO) phase at a temperature above 500~K \footnote{P. Lehuede and M. Daire, Sur la structure et les propri\'et\'es du compos\'e La$_2$CuO$_4$, C. R. Acad. Sci. Paris {\bf 276}, C-1011 (1973).}\cite{long73,gran77}, and this was soon confirmed by neutron diffraction \cite{jorg87}.  When single crystals became available, the soft phonon modes associated with tilts of the CuO$_6$ octahedra were identified by inelastic neutron scattering \cite{birg87,boni88}, 

Neutron diffraction also led to the discovery of antiferromagnetic order in LCO \cite{yama87b}; however, because of the small size of the Cu moment and the reduction of the order parameter by quantum fluctuations \cite{shir87,chak88}, it took further work to properly characterize the magnetic order \cite{vakn87,mits87}.  Despite the strong two-dimensional character of the spin correlations, the magnetic susceptibility exhibits a peak at the N\'eel temperature, $T_{\rm N}$, that turns out to be a consequence of spin-orbit-coupling effects [Dzyaloshinsky-Moriya (DM) interaction and pseudo-dipolar coupling] combined with the octahedral tilts \cite{shek92,thio94}.  The magnetic moments tend to lie within the CuO$_2$ planes, but there is a very small canting along the $c$ axis (perpendicular to the planes) that has a unique direction within each layer, but which cancels out due to an antiferromagnetic stacking of layers in the ordered state.  Applying a magnetic field of sufficient strength perpendicular to the planes causes a transition of the magnetic order to a weakly-ferromagnetic phase \cite{kast88,reeh06}.

The early samples of LCO, grown in air, exhibited semiconducting behavior, had $T_{\rm N}\approx240$~K \cite{thio88}, and occasionally showed signs of superconductivity \cite{gree87}.  Further investigation demonstrated that this was a consequence of intercalation by excess oxygen \cite{john87,sayl89}, with phase separation occurring between the antiferromagnetic and superconducting phases \cite{jorg88,hamm90,well96,well97}.  Samples on the insulating edge of the miscibility gap have the low $T_{\rm N}$, but annealing in Ar gas can raise the transition to 325~K \cite{keim92a}.  Optical measurements on such samples demonstrated the substantial charge-transfer gap of the undoped compound \cite{thio90,toku90}.

The transition to the orthorhombic phase generally results in strong twinning. With the preparation of detwinned crystals, it was discovered that the magnetic susceptibility peak at $T_{\rm N}$ seen with the field along $c$ is also present when the field is along $b$, the in-plane direction along which the octahedra tilt \cite{lavr01}.  This observation led to a reanalysis of the theoretical model for the susceptibility \cite{tabu05,silv06}.  It also motivated a reinvestigation of the crystal structure, which resulted in the discovery that the symmetry of the LTO phase is actually monoclinic \cite{reeh06}, as revealed by the appearance of weak diffraction peaks not allowed by the orthorhombic model.

More recently, a new surprise has been discovered.  Grissonnanche {\it et al.} \cite{gris19} reported the observation of a large thermal Hall effect in \lco\ and in related Sr-doped samples.  (This is a measurement in which a temperature gradient is applied along the longitudinal direction, a transverse magnetic field is applied, and the resulting temperature difference across the other transverse direction is measured.)  A follow on study \cite{gris20} found similar responses both within the CuO$_2$ planes and perpendicular to them, leading to the conclusion that the transport response must be due to phonons.  The nature of those ``chiral'' phonons remains an open question.

With this motivation, we have taken another look at the lattice symmetry and fluctuations in LCO using neutron scattering.  While confirming the superlattice peaks previously attributed to a monoclinic phase \cite{reeh06}, we find finite intensity at another set of peaks [of the type previously detected in \lsco\ (LSCO) with $x=0.07$ \cite{jaco15}] that imply a further reduction in symmetry.  These peaks correspond to the second octahedral tilt mode (orthogonal to the one that goes soft at the transition to the nominal orthorhombic phase) associated with the residual soft phonon mode found in LSCO \cite{axe89,kimu00,waki04b}.  If this mode were to go completely soft, it would result in the low-temperature-tetragonal (LTT) phase found in \lbco\ and in Nd-doped and Eu-doped LSCO \cite{axe94}.  Intriguingly, the LTT phase of the latter compounds exhibits a weak ferromagnetic response that mimics the high-field response of LCO \cite{craw93,huck04}.  It is known that the impact of spin-orbit coupling on the magnetic structure depends on the crystal electric field; therefore, the magnetic response is sensitive to the pattern of the octahedral tilts \cite{bone92,kosh94}.  Hence, we propose that the observed soft mode corresponds to the ``chiral'' phonon inferred from the thermal Hall effect \cite{gris20}.

Our measurements also capture the antiferromagnetic spin waves.  We check for a possible interaction between the spin waves and a low-energy optical mode of the type seen in Ba- and Sr-doped LCO \cite{wagm15,wagm16}, but find no evidence for such an interaction in the commensurate antiferromagnet, consistent with another recent neutron study \cite{ikeu21}.

The rest of the paper is organized as follows.  The experimental methods and initial sample characterization are described in the next section.  The main neutron scattering results and analysis are presented in Sec.~III.  We conclude with a discussion and summary of our results in Sec.~IV.

\section{Experimental Details\label{ED}}

\subsection{Crystal growth and annealing}

\begin{figure}[t]
	\centering
	\includegraphics[width=0.7\columnwidth]{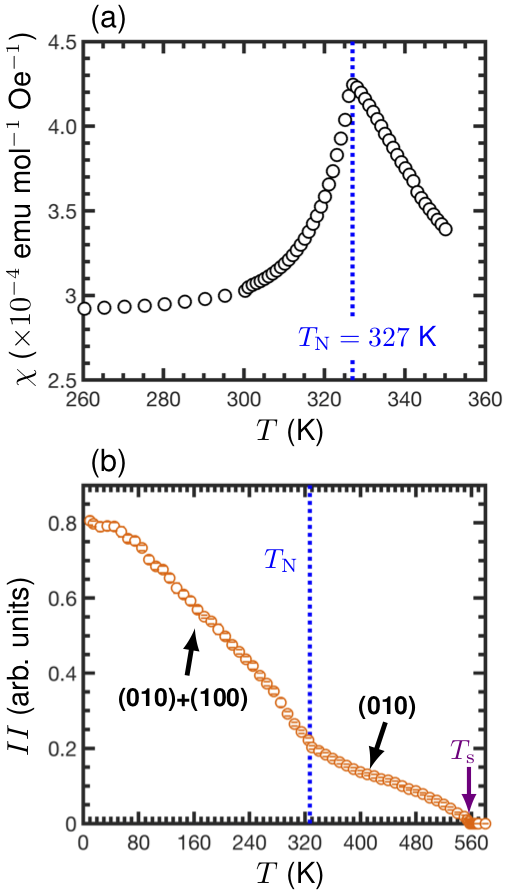}
	\caption{(a) Temperature dependence of magnetic susceptibility ($\chi =M/H$) measured with a field of $\mu_\mathrm{o}H = 0.1$~T along a random direction. Peak in the figure corresponds to N\'eel temperature $T_\mathrm{N}$. (b) Temperature dependence of the integrated intensity of $(1, 0, 0)$ magnetic and $(0,1,0)$ nuclear Bragg peaks (in close proximity due to twinning, but not resolved), illustrating the evolution of the antiferromagnetic order and the structural transition temperature, $T_{\rm s}=555\pm5$~K, to the HTT phase.}
	\label{MND}
\end{figure}

Single crystals of \lco\ (LCO) were grown at Brookhaven National Laboratory in an infrared image furnace by the traveling--solvent floating--zone method.  A large crystal of LCO was annealed in a vacuum at $10^{-3}$ torr with the following temperature profile: heat from 30~$^\circ$C to 910~$^\circ$C over 3~h, hold for 1~h, cool to 600~$^\circ$C over 12~h, cool to 500~$^\circ$C for 200 h, and finally cool to room temperature over 12 h.  The magnetic transition was determined from a measurement of the temperature dependence of magnetization using a commercial SQUID (superconducting quantum interference device) magnetometer on a small piece of the crystal.  The resulting magnetic susceptibility is shown in Fig.~\ref{MND}(a).  The peak indicates $T_{\rm N} = 327$~K, consistent with previous work \cite{keim92a}.

\subsection{Neutron scattering}

To discuss the scattering measurements, we will express the momentum transfer $\textbf{Q} = (H, K,L)$ in reciprocal lattice units (rlu) given by $(2\pi/a,2\pi/b,2\pi/c)$ with $a\approx b \approx 5.37$~\AA\ and $c\approx13.1$~\AA.  Initial neutron diffraction measurements were done on a small piece of crystal ($\sim1$~mm$^3$) from a separately annealed larger crystal.  (That larger crystal was not slowly cooled through the transition from the HTT phase, and the resulting strains eventually, after several months, caused it to shatter, providing many smaller crystals.)  Those measurements were performed at the Dimensional Extreme Magnetic Neutron Diffractometer (DEMAND, HB-3A), at the High Flux Isotope Reactor, Oak Ridge National Laboratory (ORNL). Figure~\ref{MND}(b) shows the temperature dependence of the intensity measured for the (100) antiferromagnetic reflection and the (010) monoclinic structural peak; because of twinning and finite momentum resolution, we integrated over both reflections.  The drop in intensity at $T_{\rm N}$ is consistent with the expected disappearance of the (100) intensity.  Above that, the temperature dependence of the (010) intensity indicates that the transition to the HTT phase occurs at $555 \pm 5$~K.

Inelastic neutron scattering measurements were performed on a single crystal of $11$~g at the HYSPEC spectrometer at the Spallation Neutron Source (SNS) at Oak Ridge National Laboratory.  The crystal was aligned with $(H, 0, 0)$ and $(0, K, 0)$ in the horizontal scattering plane; the vertical direction corresponds to $[0, 0, L]$. A closed-cycle He cryostat with a temperature range of 35--750~K was used to control the sample temperature. The measurements were carried out with an incident energy of $E_\mathrm{i} = 27$~meV at $T = 35$, 332, 420, and 500~K.  At each temperature, data were collected as the sample was rotated about its vertical axis in $1\degree$ steps over a range of $\sim 155 \degree$.  For all two dimensional (2D) slices and 1D cuts, shown below, the histogrammed intensities were obtained from the event data using {\tt MANTID} \cite{mantid14} and then scaled by a factor of $10^3$.

\subsection{Phonon calculations}

Density functional theory (DFT) calculations were performed using the projector augmented wave (PAW) method \cite{kresse1999ultrasoft,blochl1994projector} as implemented in the Vienna Ab-initio Simulation Package ({\tt VASP}) \cite{kresse1996efficiency,kresse1996efficient,kresse1993ab}. The standard PAW datasets included with {\tt VASP} were used to approximate the core states of all atoms. Exchange-correlation effects were treated in the local-density approximation \cite{perdew1981self} with the Hubbard-$U$ correction applied to the Cu $3d$ orbitals using the method proposed by Dudarev {\it et al.} \cite{dudarev1998electron}. We chose $U = 6$ eV as this value predicts a 1.4-eV bandgap and $0.53~\mu_{\rm B}$ magnetic moments, in good agreement with experiment \cite{tran21a}, and accurately reproduces the dispersions of the bond-stretching phonons in undoped La$_2$CuO$_4$ \footnote{T. C. Sterling, A. Holder, and D. Reznik, ``Flat bond-stretching LO phonons in undoped La$_2$CuO$_4$ calculated using LDA$+U$,'' (unpublished).}. Magnetism was assumed to be collinear and spin-orbit coupling effects were neglected. The plane-wave energy cutoff was set to 650 eV and total energy was required to converge to less that $1\times10^{-5}$~eV. During the relaxation and ground state calculations, Brillouin-zone integrations were performed using a $12\times12\times6$ $\Gamma$-centered $k$-point mesh, and energy levels were smeared with a gaussian function with width $\sigma=0.01$~eV to aid convergence (we chose a small $\sigma$ consistent with the insulating ground state). Lattice parameters were fixed at the experimental values and atomic positions were relaxed until the forces were less than 0.2~meV/\AA\ on all atoms.

To determine phonon energies and eigenvectors, we used the supercell approach as implemented in the code {\tt PHONOPY} \cite{phonopy}. We checked convergence between $2\times2\times1$ and $3\times3\times1$ supercells. The zone-center and zone-boundary energies are nearly identical between the two supercell sizes, so we concluded that the $3\times3\times1$ supercell is sufficiently large to accurately determine the force contants. We included magnetic ordering when determining the set of irreducible atomic displacements to generate the distorted supercells. {\tt VASP} was used to calculate the forces in the distorted supercells and the changes in the forces were used to approximate the force constants and populate the dynamical matrices. {\tt PHONOPY} was then used to solve the lattice dynamical equations for the energies and eigenvectors. 

We used the phonon eigenvectors and energies from the calculations to compute the inelastic neutron scattering dynamic structure factors, $S({\bf Q},\omega)$, using the {\tt SNAXS} software \footnote{D. Parshall, ``Simulating Neutron And X-ray Scans,'' https://github.com/danparshall/snaxs}. Since the computed $S({\bf Q},\omega)$ are delta functions in energy, we broadened them with a gaussian function using a full-width-half-maximum (FWHM) of 2.5~meV to approximate the effects of experimental resolution. We integrated $S({\bf Q},\omega)$ across the same ranges as the experimental binning and evaluated it at the temperatures of the experiment.

\section{Results and Analysis \label{RnD}}

\subsection{Superstructure elastic peaks and reduced symmetry\label{SEP}}

\begin{table}[b]
	\caption{Conditions for reflections within the $hk0$ plane for several space groups potentially relevant to the low-temperature phase in LCO, using the notation of Ref.~\cite{crys95}. {\newr (Application of the reflection conditions to our crystal must take account of twinning.)}
	\label{RC}}.
	\begin{ruledtabular} 
		\begin{tabular}{ccrccc}
		Acronym & \multicolumn{2}{c}{Space Group} & \multicolumn{3}{c}{Reflection conditions} \\
		& Symbol & No. & $hk0$ & $h00$ & $0k0$ \\
			\hline
			LTO & $Bmab$ & 64 & $h, k = 2n$ & $h=2n$ & $k=2n$ \\
			LTLO & $Pccn$ & 56 & $h+k=2n$ & $h=2n$ & $k=2n$ \\
			LTM1 & $Bm11$ & 8 & $h = 2n$ & $h=2n$ &  \\
			LTM2 & $P2_111$ & 4 & & $h=2n$ &  \\
		\end{tabular}
	\end{ruledtabular}
\end{table}

\begin{figure}[t]
	\centering
	\includegraphics[width=\columnwidth]{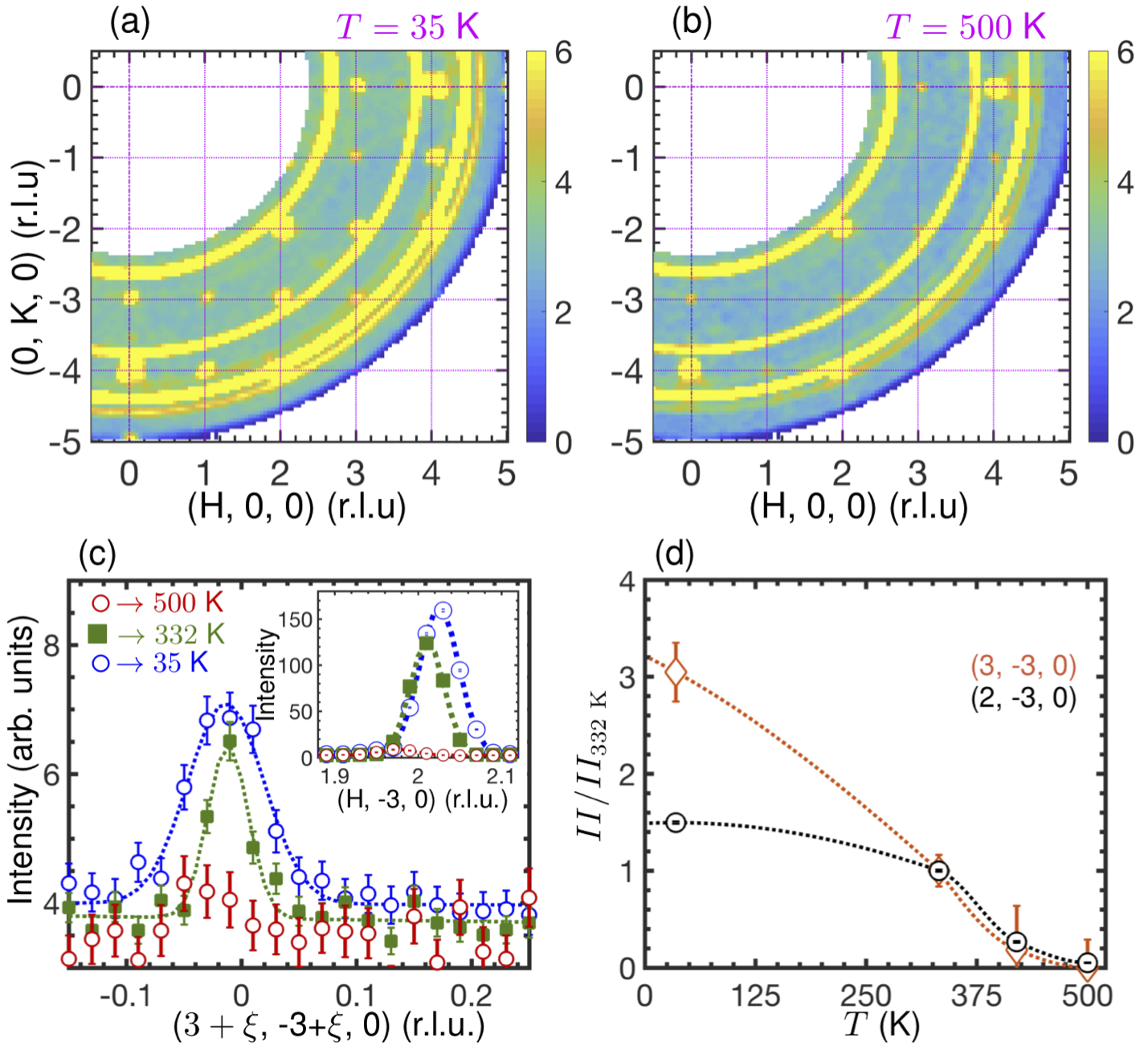}
	\caption{Slices of elastic scattering (integrated over $\pm0.3$~rlu in $L$) at temperatures (a) 35~K,  (b) 500~K. (c) Transverse cuts through $(3,-3,0)$ [inset: $(2,-3,0)$] at $T=35$, 332, and 500~K; dashed lines are gaussian fits.  (d) Temperature dependence of the integrated intensity for $(3, -3, 0)$ and $(2, -3, 0)$ peaks, normalized at 332~K; dotted lines are shape-preserving interpolations.  {\newr The continuous rings are diffraction from the Al sample holder.}}
	\label{E0}
\end{figure}

The deviations of the lattice symmetry from the LTO phase are apparent from a study of diffraction peaks in the $(H,K,0)$ plane.  As indicated in Table~\ref{RC}, the only allowed peaks within $Bmab$ require that $H$ and $K$ each be even.  Figure~\ref{E0}(a) shows elastic scattering over a regime of moderate $Q$ obtained at 35~K; one can clearly see many peaks that are inconsistent with $Bmab$.  A complication at this temperature is that we have antiferromagnetism and associated Bragg peaks.  

\begin{figure}[b]
	\centering
	\includegraphics[width=0.9\columnwidth]{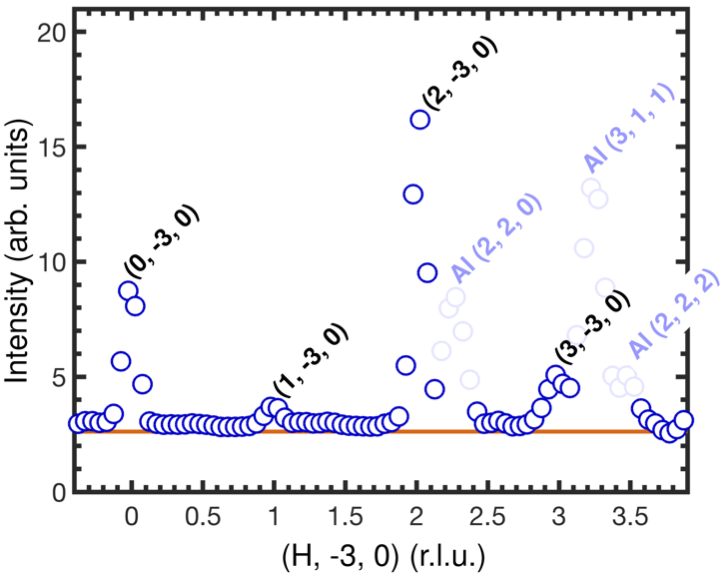}
	\caption{Elastic scattering along $(H,-3,0)$.  Diffraction peaks from Al sample holder are faded; red line indicates the background level.}
	\label{E1}
\end{figure}

To further highlight this, Fig.~\ref{E1} shows a cut along $(H,-3,0)$ obtained at 332~K, above $T_{\rm N}$.  None of the peaks observed here are allowed in LTO.   The stronger peaks, at $(0,-3,0)$ and $(2,-3,0)$, are consistent with the monoclinic space group $Bm11$, as proposed by Reehuis {\it et al.} \cite{reeh06}; we will label this phase as low-temperature monoclinic 1 (LTM1). 
{\newr [Please note: Reehuis {\it et al.} \cite{reeh06} measured a detwinned sample, so they were able to identify that there is intensity for peaks at $(0,K,0)$ with $K$ odd but not for $(H,0,0)$ with $H$ odd.  In contrast, our crystal is twinned, so that, for example, we see peaks at both $(0,-3,0)$ and $(3,0,0)$.]}
 In addition, there are also peaks at $(1,-3,0)$ and $(3,-3,0)$, that satisfy $H+K=2n$; in the absence of the monoclinic reflections, these would be consistent with the low-temperature less-orthorhombic (LTLO) phase observed in the phase diagram of \lbco\ \cite{huck11}.    To explain the presence of both sets of peaks in a single phase, we propose space group $P2_111$ and label the phase LTM2.

Figure~\ref{E0}(c) and (d) show the $T$ dependence of the $(3,-3,0)$ and $(2,-3,0)$ peaks.  They both show a large fall off in intensity around 400~K, and are difficult to detect at 500 K.  Given that we have already seen in Fig.~\ref{MND}(b) that the (010) monoclinic peak survives to $T_{\rm s}$, it seems likely that the drop in intensity of the peaks at moderate $Q$ is due to the Debye-Waller factor.  {\newr We suspect that there is a single transition at $T_s$ from the LTM2 phase to the HTT phase; however, further measurements would be necessary to confirm this.}  It is certainly the case that there are lots of fluctuations associated with the LTLO peaks, as we will discuss next.

\subsection{Soft tilt modes\label{SM}}

\begin{figure}[b]
	\centering
	\includegraphics[width=\columnwidth]{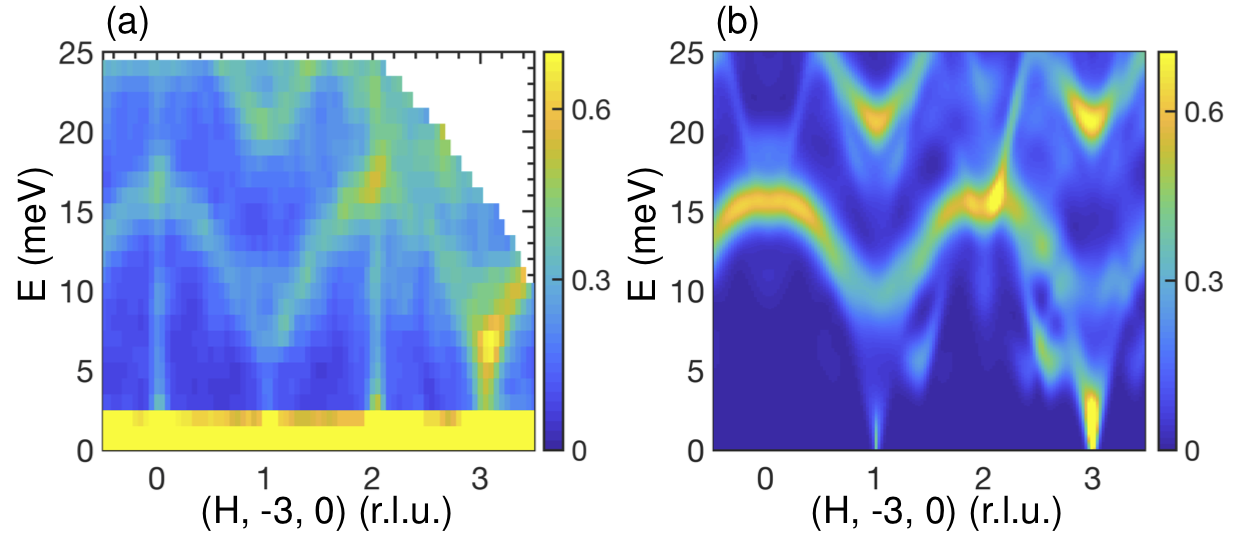}
	\caption{(a) Inelastic neutron scattering spectra at $332$~K illustrating dispersion along $[H, 0, 0]$ at $K = -3$ (integrated over $\pm0.1$ in $K$ and $\pm0.3$ in $L$). One can see soft-mode phonons at $H=1$ and 3, but only spin waves at $H=0$ and 2, with  the acoustic phonons from the corresponding monoclinic peaks being too weak to detect. (b) Calculated phonons for the same range, assuming LTT structure.}
	\label{MvS}
\end{figure}

Figure~\ref{MvS}(a) shows a slice of inelastic scattering along the same range of ${\bf Q}$ (and the same temperature) as used for elastic scattering in Fig.~\ref{E1}.  Here we see diffuse phonons rising from the points $H=1$ and 3, but nothing except vertically-dispersing spin excitations at $H=0$ and 2.   The calculated phonons, assuming an LTT structure, are shown in Fig.~\ref{MvS}(b).  (Note that a single scale factor has been used for the calculated intensities for all of the comparisons to be shown, including phonons around strong Bragg peaks.)  The calculation shows relative weights of acoustic phonons at $H=1$ and 3 that are similar to the diffuse response in the measured spectra.  
 
The intensity of acoustic modes must scale with the intensity of the associated Bragg peak.  The pattern that we observe between experiment and calculations suggests that if the sample transformed to the LTT phase, the $(1,-3,0)$ and $(3,-3,0)$ peaks would be much stronger than the LTM1 peaks at $(0,-3,0)$ and $(2,-3,0)$.  In reality, what we find is that much of the measured phonon weight associated with LTLO wave vectors is peaked above 5~meV, consistent with soft-mode behavior.

\begin{figure}[t]
	\centering
	\includegraphics[width=\columnwidth]{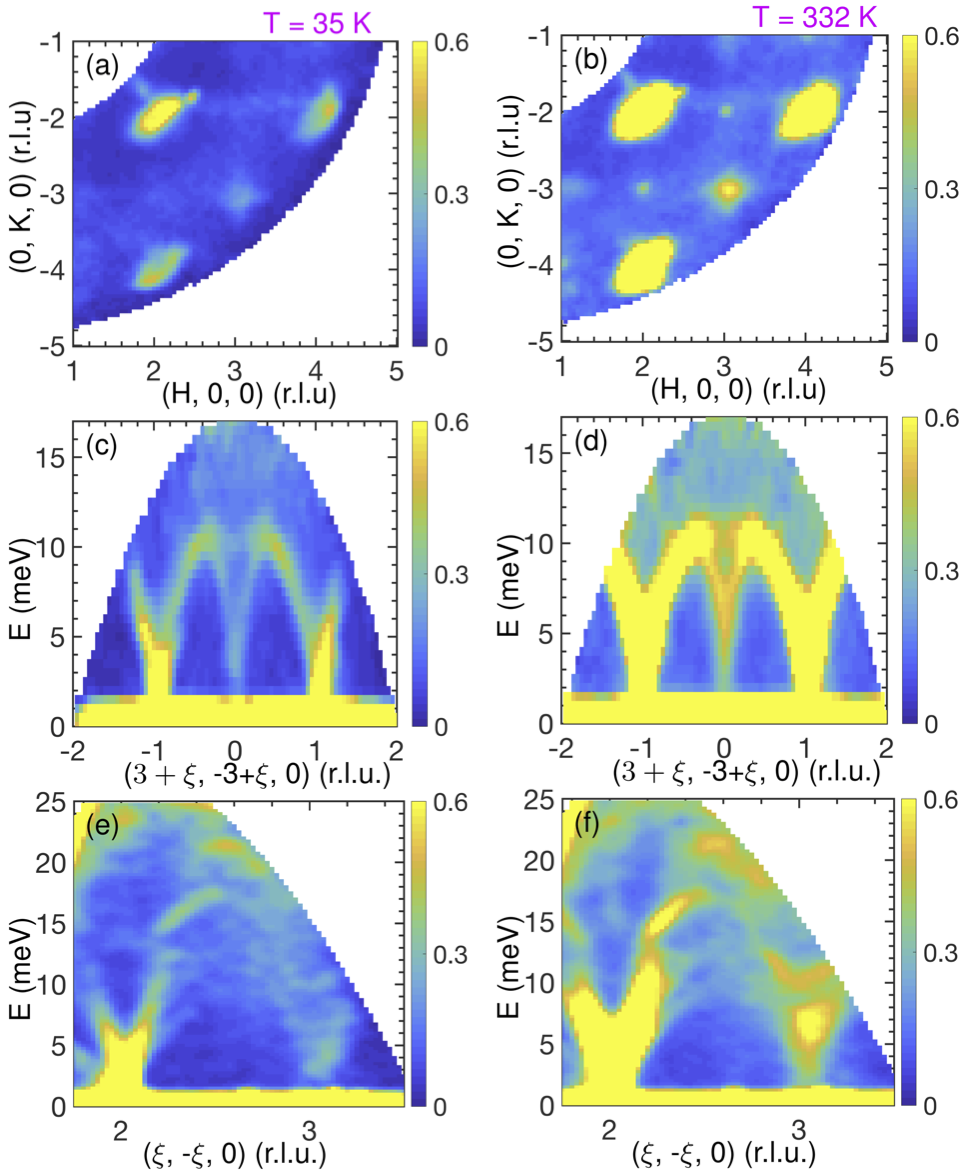}
	\caption{Comparisons of phonons in the vicinity of $(3,-3,0)$.  Constant-energy slices at $E=5$~meV, integrated over $\pm2$~meV in $E$ and $\pm0.5$~rlu in $L$ for (a) 35~K, (b) 332~K.  Transverse dispersion for (c) 35~K, (d) 332~K. Longitudinal dispersion for (e) 35~K, (f) 332~K.  The latter 4 plots were integrated over $\pm0.5$ rlu in $L$ and $\pm0.1$ rlu in the in-plane direction transverse to each plot.}
	\label{NMSM}
\end{figure}

\begin{figure}[t]
	\centering
	\includegraphics[width=\columnwidth]{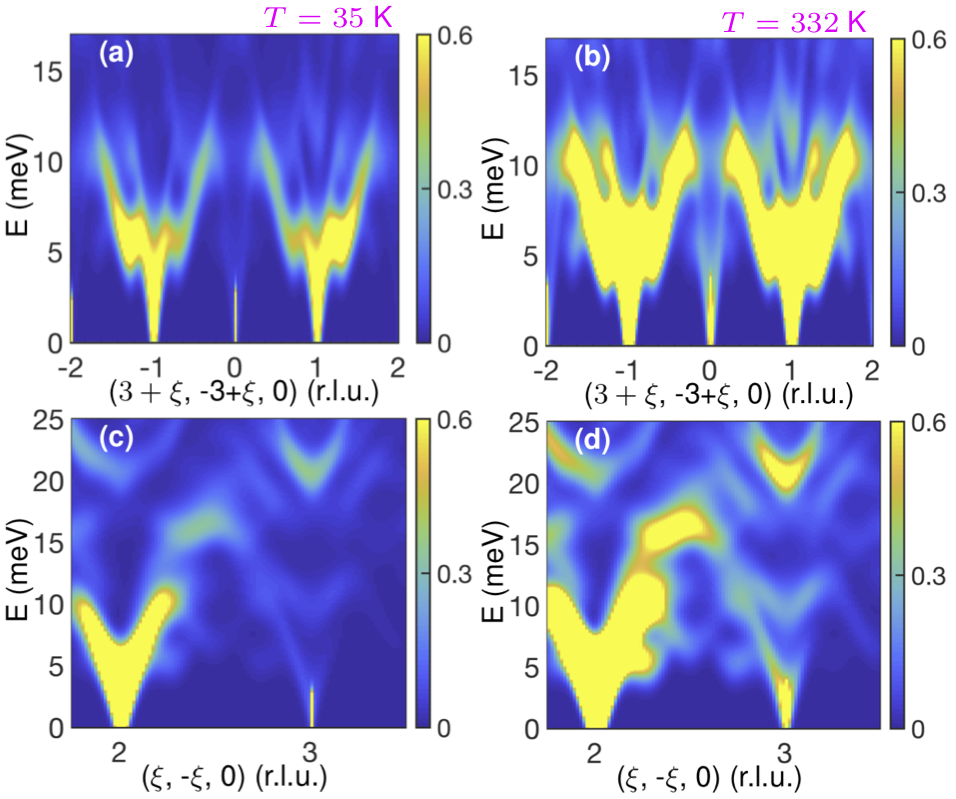}
	\caption{Calculated phonons in the vicinity of $(3,-3,0)$ for comparison with data in Fig.~\ref{NMSM}.    Transverse dispersion for (a) 35~K, (b) 332~K. Longitudinal dispersion for (c) 35~K, (d) 332~K.}
	\label{NMSMc}
\end{figure}

We now take a more systematic look at the LTLO lattice fluctuations, focusing on the $(3,-3,0)$ wave vector.  Figure~\ref{NMSM}(a) and (b) show constant-energy slices centered at $E=5$~meV and plotted in the $(H,K,0)$ plane for two temperatures.  The acoustic phonons from neighboring fundamental Bragg points have an elliptical shape, reflecting the anisotropic velocities for longitudinal (L) and transverse (T) phonons.  [Note that the neutron cross section is proportional to $({\bf Q}\cdot\beps)^2$ where $\beps$ is the phonon eigenvector, which results in anisotropic sensitivity to L and T modes.]  The excitations at $(3,-3,0)$ are more symmetric, consistent with the weight coming from soft phonons, rather than sharply defined L and T acoustic modes.  At 332~K, we can also see sharp spots at $(2,-3,0)$ and $(3,-2,0)$ due to damped spin fluctuations, which are not apparent at 35~K due to a significant excitation gap in the magnetically-ordered state.

The dispersions of phonons about $(3,-3,0)$ in the T and L directions are plotted in Fig.~\ref{NMSM}(c), (d) and (e), (f), respectively.  The T mode appears to connect continuously with the acoustic (A) modes from neighboring reciprocal lattice points (although the calculated dispersions suggest some small gaps from anticrossing modes at 8 meV and above).  The {\newr LA} excitations {\newr disperse linearly with strong intensity up to $\sim 10$~meV, where the intensity suddenly falls off, with intensity reappearing near 15~meV.  This behavior can be understood as} an avoided crossing {\newr(due to hybridization)} with an optical mode at $\sim10$~meV (due to a mode involving La motion), so the weight of the branch connecting to the LA mode at $(2,-2,0)$ is weak.  With increasing $T$, the soft-mode weight shifts upward in energy.

For comparison, calculated neutron scattering from phonons is presented in Fig.~\ref{NMSMc}.  There is good qualitative consistency between the measured and calculated phonons.  One feature that sticks out in the calculated T dispersions is the dip in the acoustic dispersion near 5~meV; inspection of the eigenvector indicates that this feature is associated with La and O motion, with almost no participation from the Cu sublattice.  There are hints of related behavior in the 35-K data, but the transverse $Q$ resolution is limited by the orthorhombic twinning that effectively increases the sample mosaic.

\begin{figure}[t]
	\centering
	\includegraphics[width=\columnwidth]{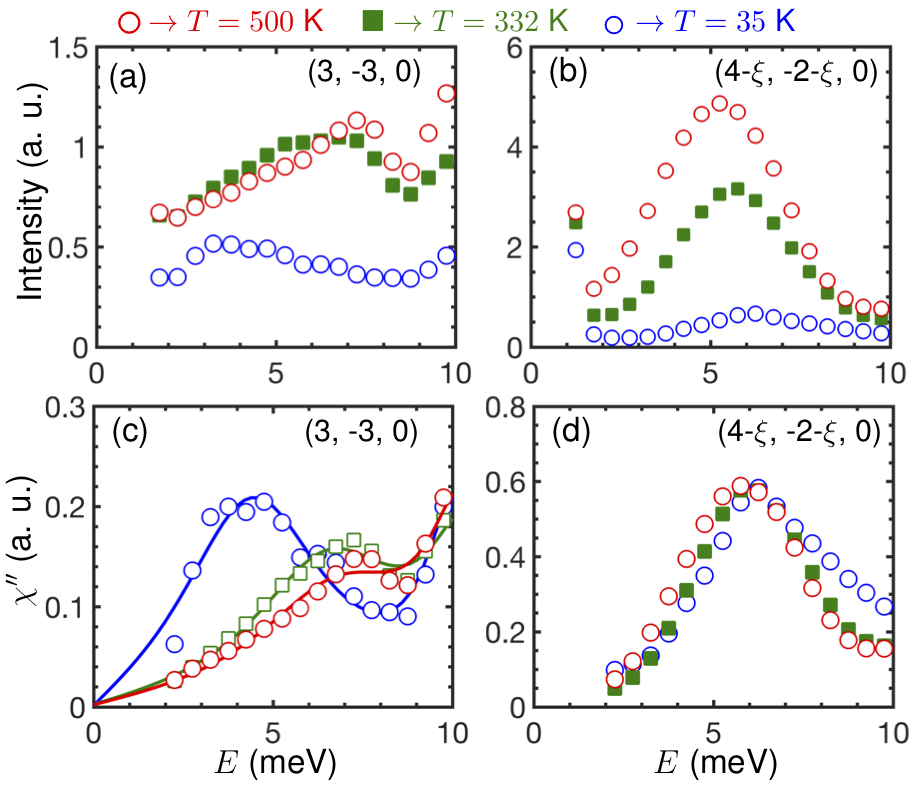}
	\caption{Intensity of phonon scattering for several temperatures 
	{\newr in the vicinity of} (a) $(3,-3,0)$, (b) $(4-\xi,-2-\xi,0)$ with $\xi=0.225$.  In each case, the intensities average over windows of $\pm0.1$ rlu in the $[1,-1,0]$ direction, $\pm0.05$ in the $[1,1,0]$ direction, and $\pm0.5$ in $L$.  The corresponding results for $\chi''$ are plotted in (c) and (d), respectively.}
	\label{DB}
\end{figure}

{\newr  The temperature dependence of the phonons at $(3,-3,0)$ is anomalous, and to demonstrate this, we will compare with the behavior of TA phonons dispersing from a neighboring Bragg point.  Considering Fig.~\ref{NMSM}(c), we pick the points corresponding to $\xi=0$ and 0.775.  The latter point can be described more conventionally as $(4-\xi',-2-\xi',0)$ with $\xi'=0.225$.}
Figure~\ref{DB}(a) and (b) compare the temperature dependence of phonon intensities 
{\newr in the vicinity of these two points}.  For the latter, we see that the scattered intensity grows with temperature.  If we divide out the detailed balance factor, we obtain the imaginary part of the dynamic susceptibility, $\chi''({\bf Q},E)$, plotted in Fig.~\ref{DB}(d), which is almost independent of $T${\newr, as is typical for conventional phonons}.  The situation is different for $(3,-3,0)$: the intensity changes much less with $T$ [Fig.~\ref{DB}(a)], while $\chi''$ changes quite a bit [Fig.~\ref{DB}(c)].  

To quantify the behavior of the soft mode with temperature, we fit the energy dependence shown in Fig.~\ref{DB} (c) using the damped harmonic oscillator (DHO) formula:
\begin{equation}\label{DHO}
\chi^{\prime \prime}(\textbf{Q}, E) = A \times \dfrac{\gamma E E_{\bf{q}}}{\big[E^2 \minus E_{\bf q}^2]^2 + (\gamma E).^2} 
\end{equation}
where $E_{\bf q}^2 = \Delta_p^2+v^2q^2$, ${\bf q} = {\bf Q} - (3, -3, 0)$, $\Delta_p$ is an energy gap, $v$ is the average acoustic velocity, the damping factor $\gamma$ corresponds to the FWHM,  and $A$ is the amplitude. The fitted values are listed in Table~\ref{TM}.  
At all temperatures, the soft phonon is under-damped, with $\Delta_p > \gamma/2$. 
The increase of the soft mode gap with temperature is consistent with the soft-mode behavior in a perovskite structure \cite{tkac21}. 

\begin{table}[t]
	\caption{DHO fitting parameters, corresponding to Eq.~\ref{DHO}, for fits to phonon $\chi''$ at $(3,-3,0)$ shown in Fig.~\ref{DB}(c).  Uncertainties correspond to one standard deviation. \label{TM}}.
	\begin{ruledtabular} 
		\begin{tabular}{cccc}
			$T$~(K) & $\Delta_p$~(meV) & $\gamma$~(meV) & $A$~(arb. units)\\
			\hline
			$35$ & $4.6 \pm 0.1$ & $4.7 \pm 0.4$ & $1.8 \pm 0.1$\\
			 $332$ & $6.7 \pm 0.2$ & $5.1 \pm 0.5$ & $1.4 \pm 0.2$\\
			$500$ & $7.1 \pm 0.1$ & $5.7 \pm 0.8$ & $1.3 \pm 0.2$\\
		\end{tabular}
	\end{ruledtabular}
\end{table}

\subsection{Spin Fluctuations\label{SF}}

In studies of antiferromagnetic spin fluctuations in \lbco\ and \lsco, Wagman {\it et al.} \cite{wagm15,wagm16} have identified evidence for spin-phonon coupling at energies in the range of 15-20~meV.  We have observed a similar effect in our own samples of \lbco\ \cite{xu14} and \lsco\ \cite{li18}, and even in superconducting La$_{2-x}$Ca$_{1+x}$Cu$_2$O$_6$ \cite{schn19}.  In most of these cases, the low-energy magnetic excitations have an incommensurate modulation, and it is possible that the coupling to phonons is related to that spatial variation.  If that is the case, then the effect should be absent in the case of commensurate antiferromagnetism.  We decided to test this in our sample.

\begin{figure}[t]
	\centering
	\includegraphics[width=\columnwidth]{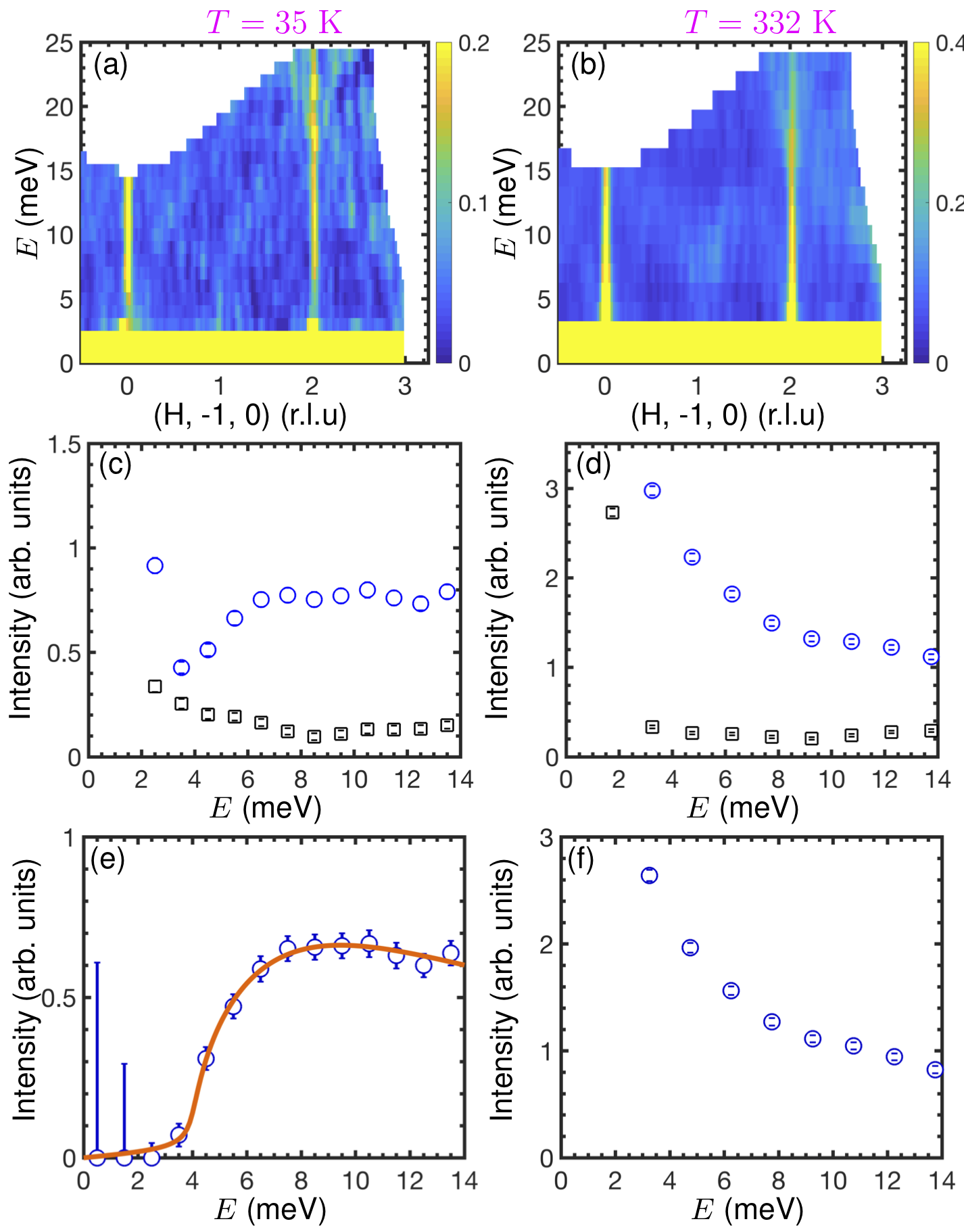}
	\caption{(a) and (b) Inelastic neutron scattering spectra showing spin fluctuations dispersion in two magnetic Brillouin zones for temperatures below and above $T_\mathrm{N}$, respectively. Slices are created after averaging along $[0, K, 0]$ and $[0, 0, L]$ by $\pm 0.08$ and $\pm 0.3$ r.l.u., respectively. (c) and (d) 1D energy cuts at $(0, -1, 0)$ showing the energy dependence of the spin fluctuations, blue circles, for $T_\mathrm{N}$, respectively. Energy cuts for the spin fluctuations are obtained after integrating the data in $H$ by $\pm 0.05$. Black open squares are the background obtained at the wavevectors (0.15 $\pm$ 0.05, -1, 0). (e) Energy cuts obtained from (c) after subtracting the background and Gaussian lineshape fits to the zero energy signal and highlights the presence of the spin-gap at 35~K$ < T_\mathrm{N}$. The solid orange line is fit to the spin gap equation in Ref.~\cite{rama17}. (f) Energy cuts obtained after subtracting the background in (d).}
	\label{SFSG}
\end{figure}

Figure~\ref{SFSG}(a) shows inelastic scattering measured along ${\bf Q}=(H,-1,0)$ at 35~K, where one can see the spin waves dispersing vertically at $H=0$ and 2.  Similar results are obtained at 332~K, as shown in Fig.~\ref{SFSG}(b).  From previous work \cite{keim93}, there should be a spin gap for out-of-plane excitations of 5 meV.  For our geometry, the contribution from the in-plane fluctuations, with spin gap of 2.5~meV, should be only 1/3 of the signal.  To compare our 35-K data centered at $(0,-1,0)$ with this expectation, we have integrated the intensity over {\bf Q}, and plot the resulting intensity vs.\ $E$ in Fig.~\ref{SFSG}(c), together with background determined at a nearby point.  The net magnetic signal, after subtracting the background and gaussian peak fit to the elastic magnetic peak, is plotted in Fig.~\ref{SFSG}(e).  As a quantitative check of the spin gap, we have fit the energy dependence using a lorentizian formula applied in previous studies \cite{rama17,lake99}; we obtain a gap $\Delta_s=4.0\pm0.1$~meV, which is very close to the weighted average (4.17~meV) of the previous results cited above \footnote{The other parameters determined in the fit are an overall damping factor $\Gamma=8.6\pm0.4$~meV for the lorentzian function, and a damping factor for the gap edge $\Gamma_s=0.2\pm0.1$~meV.}.
In comparison, heating above $T_{\rm N}$ converts the static order into damped spin fluctuations, filling in the spin gap, as illustrated in Fig.~\ref{SFSG}(d) and (f).

\begin{figure}[t]
	\centering
	\includegraphics[width=\columnwidth]{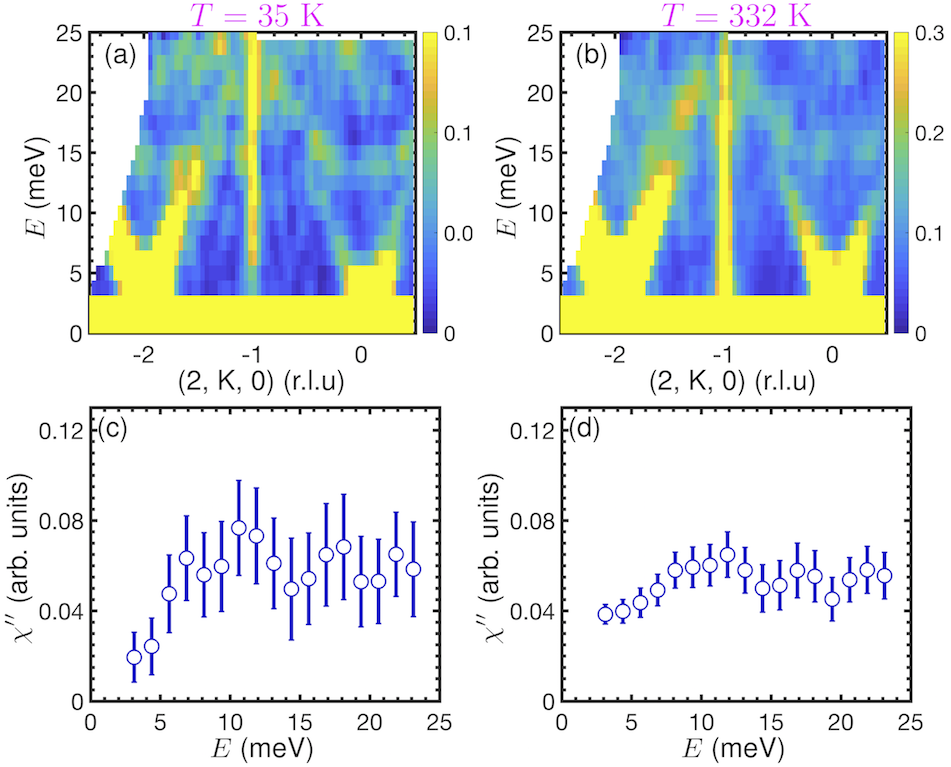}
	\caption{(a) and (b) Inelastic neutron scattering spectra for $35$ and $332$~K, respectively, illustrating the energy dependence of spin fluctuation dispersions along $[0, K, 0]$.  Data were averaged over $\pm0.05$ rlu in $H$ and $\pm0.3$ rlu in $L$. (c) and (d) $\chi^{\prime\prime}(E)$ for $35$ and $332$~K, respectively, obtained after subtracting the background from intensity vs $E$ plot for spin fluctuations and then applying the bose-factor correction. Intensity vs $E$ plot is a 1D cut at the magnetic zone center $(2, -1, 0)$ and the background is similar plot taken away from magnetic zone center.}
	\label{NESF}
\end{figure}

To look for evidence of spin-phonon coupling, we must shift to the magnetic signal at $(2,-1,0)$, where we can transfer enough energy to reach the energy range of interest.  Figure~\ref{NESF}(a) and (b) show slices of the magnon and phonon dispersions along $(2,K,0)$ at two temperatures, where one can see the phonons that disperse towards the magnetic {\bf Q} from the neighboring Bragg points.  To examine the energy dependence of the spin waves, we integrated the excitations in the window $H=2\pm0.05$, $K=-1\pm0.075$, and subtracted averaged background from the windows $K=-1.1\pm0.02$ and $K=-0.9\pm0.02$, both with $H=2\pm0.05$.  The resulting signal is plotted in the form of {\bf Q}-integrated $\chi''(E)$ in Fig.~\ref{NESF}(c) and (d).  As one can see, the signal for $E\gtrsim8$~meV is essentially constant, consistent with expectations for {\bf Q}-integrated 2D spin waves.  There is no sign of any significant interaction with phonons for this undoped, commensurate antiferromagnetic system in either the ordered or paramagnetic phases.

\section{Discussion and Summary}

We have provided evidence for LTLO-like reflections in LCO.  The elastic intensities are weak, but there is plenty of intensity in the soft phonons at these positions, qualitatively consistent with phonons calculated for the LTT phase.  Following on our previous observation of such peaks in LSCO $x=0.07$ \cite{jaco15}, it seems probable that this behavior is typical of the low-temperature structural phase in LSCO.  This finding is significant because the distortion results in inequivalent Cu-O bonds in the planes, which are relevant to pinning and ordering charge stripes \cite{hunt99,crof14,tham14}.

The temperature dependence of the soft-mode at $(3,-3,0)$ is clearly distinct from that observed for conventional acoustic phonons.  In the latter case, $\chi''$ is approximately constant with temperature, so that the intensity (obtained by multiplying $\chi''$ by the detailed-balance factor) grows with $T$ due to weight transferred from the associated Bragg peak.  The substantial low-temperature weight of the soft mode is consistent with LTT-like tilt fluctuations that are not going to order, representing significant entropy at low $T$.  Besides the case of LSCO $x=0.07$ \cite{jaco15}, we have seen related entropic modes in systems such as superconducting (Pb$_{1-x}$Sn$_x$)$_{1-y}$In$_y$Te \cite{ran18,sapk20} and in both superconducting and non-superconducting Fe$_{1+y}$Te$_{1-x}$Se$_x$ \cite{fobe16a}.

Of course, there are also peaks present that indicate monoclinic symmetry \cite{reeh06}, and we have proposed that the space group consistent with all observed peaks is $P2_111$.  (Note that the observed unit cell vectors are consistent with orthorhombic symmetry \cite{reeh06}; it is the set of atomic positions within the unit cell that determines the symmetry to be monoclinic.)  The LTLO-like peaks are understandable in terms of octahedral tilt instabilities \cite{pick91}, but it is also desirable to have some physical understanding of the monoclinic symmetry.  Consider first a larger distortion that is not explained by tilting of rigid octahedra.  The large LTO-like tilt is along the $b$ axis.  If this involved rigid rotations, then we would have $b<a$, contrary to experiment.  Instead, there is a shear distortion of the CuO$_2$ planes, and an early analysis showed that it is a secondary effect induced by La-La and La-apical-oxygen interactions \cite{pive91}.  (This effect is in some ways similar to the shear distortion that leads to the monoclinic phase of Fe$_{1+y}$Te, ``parent'' compound of another superconductor family \cite{zali12}.)  A significant bonding between apical-O and La atoms has also been found in a new DFT-based study that concludes the ground state to be LTLO \cite{lee21b}. 

Another DFT-based study evaluated the phonons in the HTT phase, finding two unstable modes \cite{wang99}.  From the calculated eigenvectors of these modes, one corresponds to the LTO-like tilts, while the other involves a translation (parallel to the planes and transverse to the tilt) of the apical O ions relative to the La ions.  A static displacement of the latter type would provide an explanation for the monoclinic symmetry.

\begin{figure}[t]
	\centering
	\includegraphics[width=\columnwidth]{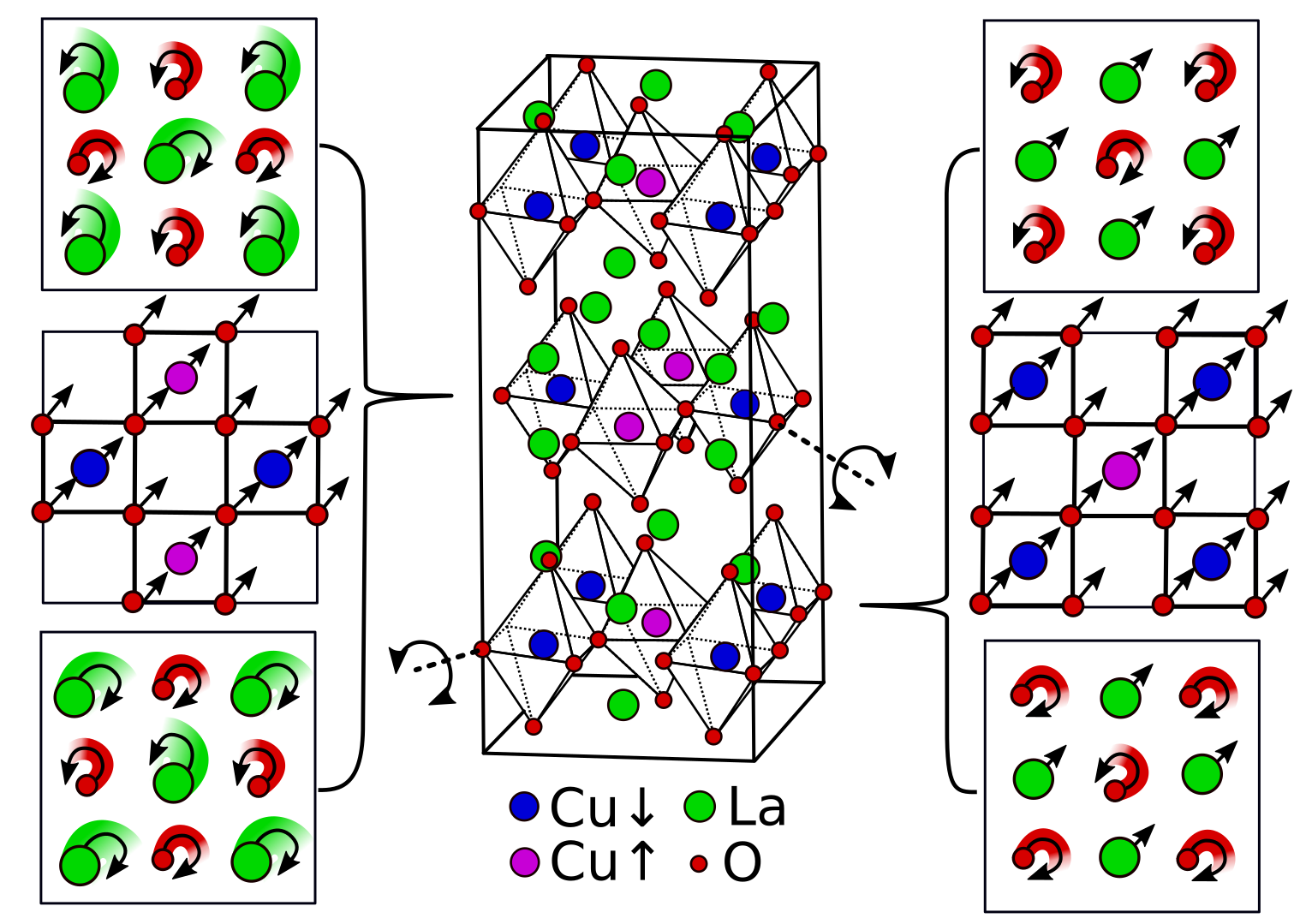}
	\caption{Schematic diagram of the LA eigenvector with ${\bf q} = [0.1,-0.1,0]$, where  $E({\bf q})\approx5$~meV.  Atomic structure of 1.5 unit cells is indicated in the middle, with La in green, O red, and Cu with antiparallel spins in blue and magenta.  Left panels indicate displacements of atoms in middle layer for the top La-O layer, Cu-O layer, and bottom La-O layer.  Right panels show same for bottom layer.}
	\label{eig}
\end{figure}

While our phonon calculations were performed on the LTT structure rather than with the experimental symmetry, an examination of the eigenvector for an LA phonon in the $[1,-1,0]$ direction, shown in Fig.~\ref{eig}, yields intriguing insights.  While the in-plane Cu and O atoms exhibit linear displacements, the apical oxygens and half of the La ions exhibit circular motions.  (The symmetry between the layers is broken by the orthogonal orientation of the octahedral tilt axes between neighboring layers, as indicated by the semicircular arrows in the middle panel.)  The rotations of atoms above and below the planes are equal and opposite.  For the experimental structure, the rotations are likely to be much more elliptical.

Turning to the magnetic response of the system, the field-induced weak ferromagnetism in LCO is due to the DM interaction, as mentioned in the introduction.  The distortions of the octahedra will impact the crystal-field splittings, and these impact the DM interaction through spin-orbit coupling.  In zero field, we already have reduced lattice symmetry and associated low-energy fluctuations.  {\newr In the weakly-ferromagnetic phase, the acoustic phonons can couple to the magnetic response, and this coupling could be relevant to understanding the recent experimental observations of the thermal Hall effect  \cite{gris19,gris20}.}


{\newr 
The thermal Hall effect involves a transverse component of the thermal conductivity that results in an anisotropic response in the presence of an orthogonal magnetic field.  In the case of LCO \cite{gris19}, which is a Mott insulator, the excitations involved cannot be mobile charge carriers, leaving phonons or magnetic excitations.  Theoretical analyses find that, to explain the thermal Hall effect, these bosonic excitations must have a topological character \cite{sait19,yang20}.   Some studies \cite{han19,sama19} have shown that magnetic excitations of distinct types that might be associated with quantum critical effects could result in an enhanced thermal Hall effect, but later experimental results lead to the conclusion that phonons must be the chief contributor \cite{gris20}.

The soft phonons in LCO may couple to the weak ferromagnetism in a $c$-axis magnetic field of sufficient strength.  We propose that they may be relevant to the experimental observations of the thermal Hall effect \cite{gris20}.  Of course, the experimental story has become more complicated, as large thermal Hall effects have now also been observed in the antiferromagnetic insulators Nd$_2$CuO$_4$ and Sr$_2$CuO$_2$Cl$_2$ \cite{boul20}.  While the Cu sites in these compounds lack apical O neighbors, they nevertheless exhibit unusual behaviors tied to magnetism.  Nd$_2$CuO$_4$ exhibits three distinct magnetic phases as a function of temperature, with Nd moments coupling to Cu in the phase that sets in below 30~K \cite{mats90}.  An optical second-harmonic-generation study of Sr$_2$CuO$_2$Cl$_2$ has provided evidence for an unexpected order parameter tied to the antiferromagnetic order and leading to reduced lattice symmetry \cite{dela21}. Hence, the possible coupling of phonons to unusual magnetic responses is not ruled out for these compounds. 

Another complication is that a large thermal Hall effect has been reported in SrTiO$_3$ \cite{li20c}, a nonmagnetic quantum paraelectric.  This has been explained in terms of an extrinsic mechanism of Berry-curvature-induced skew scattering from defects \cite{chen20}.  While there are challenges in applying this picture to cuprates, skew scattering from magnetic defects is a possibility \cite{chen20}. 

Besides LCO, a substantial thermal Hall effect has also been observed in hole-doped LSCO with $p=0.06$.}
Along these lines, it is relevant to note that Balendent {\it et al.} \cite{bale10} reported a neutron spin-flip {\newr ({\it i.e.}, a magnetic scattering)} signal at (110) [equivalent to (100) when indexed relative to the HTT unit cell] below $\sim100$~K in LSCO $x=0.085$.  That study was motivated by the search for evidence of loop currents \cite{varm06}, and the authors were not aware that there might be a nuclear peak at (110).  Their observation suggests a possible coupling of the lattice to the magnetic response, and deserves further investigation.

Finally, we have found no evidence for interaction of the antiferromagnetic spin waves with phonons in the 15-20 meV range.  Our results are consistent with another recent study that found such interactions in superconducting LSCO but not in LCO \cite{ikeu21}.  The interaction of phonons with spin stripes in hole-doped CuO$_2$ planes is an interesting problem that will need further study.

\section{Acknowledgments}

We thank Melissa Graves-Brook for assistance with the experiment at HYSPEC.  The work at Brookhaven was supported by the U. S. Department of Energy (DOE), Office of Basic Energy Sciences, Division of Materials Sciences and Engineering, under Contract No.\ DE-SC0012704.  
T.S. and D.R. were supported by the DOE, Office of Basic Energy Sciences, Office of Science, under Contract No. DE-SC0006939. A portion of this research was performed using the Eagle computer operated by the Department of Energy's Office of Energy Efficiency and Renewable Energy and located at the National Renewable Energy Laboratory.  T.R. thanks Aaron Holder for help with the calculations.  This research used resources at the High Flux Isotope Reactor and the Spallation Neutron Source, DOE Office of Science User Facilities operated by the Oak Ridge National Laboratory.

\bibliography{LNO,theory,neutrons,ref,fe_sc}


\end{document}